\documentclass[twocolumn,prd,amsmath,floatfix,preprintnumbers,nofootinbib,a4paper]{revtex4}

\usepackage{epsfig, graphicx}
\usepackage[T1]{fontenc}

\usepackage{natbib}

\begin{document}

\title{An alternative approach to the limits of predictability in human mobility}

\author{Edin Lind Ikanovic and Anders Mollgaard}

\affiliation{Niels Bohr Institute, University of Copenhagen,\\
 2100 Copenhagen, Denmark}

\begin{abstract}
Next place prediction algorithms are invaluable tools, capable of increasing the efficiency of a wide variety of tasks, ranging from reducing the spreading of diseases to better resource management in areas such as urban planning. In this work we estimate upper and lower limits on the predictability of human mobility to help assess the performance of competing algorithms. We do this using GPS traces from 604 individuals participating in a multi year long experiment, The Copenhagen Networks study. Earlier works, focusing on the prediction of a participant's whereabouts in the next time bin, have found very high upper limits ($>90\%$). We show that these upper limits are highly dependent on the choice of a spatiotemporal scales and mostly reflect stationarity, i.e. the fact that people tend to {\it not} move during small changes in time. This leads us to propose an alternative approach, which aims to predict the next location, rather than the location in the next bin. Our approach is independent of the temporal scale and introduces a natural length scale. By removing the effects of stationarity we show that the predictability of the next location is significantly lower ($\sim$ 71.1\%) than the predictability of the location in the next bin.
\end{abstract}

\maketitle

\section{Introduction}
\label{intro}
The understanding of human mobility patterns has changed greatly in the last couple of decades. This has mainly been due to new technologies enabling human displacements to be studied with higher accuracy over a longer period of time. Starting with the tracking of bank notes \cite{brockmann2006scaling} as a proxy for human movement, studies quickly evolved towards the current use of hand held devices for tracking, using either GSM data \cite{gonzalez2008understanding,Song2010}, connections to wifi hotspots \cite{qian2013impact} or GPS receivers \cite{rhee2011levy} to determine location. The main results from these studies have been the discoveries of power laws governing step size and wait time distributions \cite{brockmann2006scaling}, a universal probability density governing human mobility \cite{song2010modelling}, and simple models capturing many statistical features of human mobility \cite{song2010modelling,rhee2011levy,jiang2016timegeo,pappalardo2016modelling}. It has furthermore been explored how mobility is affected by recency \cite{barbosa2015effect}, exploration \cite{pappalardo2015returners}, and return to previously visited places \cite{song2010modelling} and friends \cite{toole2015coupling}. Such discoveries and models can help predict the spread of diseases \cite{colizza2007modeling} and cellphone viruses \cite{kleinberg2007computing}, and also enhance socio-economic forecasting \cite{gabaix2003theory,pappalardo2016analytical,frias2012relationship}, city planning \cite{makse1998modeling} and many other fields \cite{kitamura2000micro,krings2009urban,rhee2011levy} . Further contribution to progress in these areas can be made if geolocation data can be used to accurately predict an individual's future whereabouts. A crucial part of this work is the construction of viable evaluation mechanisms, thereby raising the question: what are the upper and lower limits, $\Pi^{max}$ and $\Pi^{min}$, on the predictability of human mobility? 

This question was initially investigated using call detail records from 45,000 cellphones \cite{Song2010}. Each call corresponded to a known location represented by a Voronoi cell, around the closest cell tower, with an average area of $3 $ km$^2$. The known locations were grouped into 1 hour bins, giving a history of locations $T_i$, for each user $i$. The work focused on determining how well the best possible algorithm can predict the location of an individual in the next time bin, given $T_i$. They reported an upper limit narrowly peaked at $\Pi^{max} = 93 \%$ and a lower limit of $\Pi^{min} = 70 \%$. 

This work led to questions being raised about possible biases introduced when using call detail records \cite{ranjan2012call} and about the influence of spatiotemporal scales \cite{lin2014mining}. 
The temporal resolution \cite{jensen2010estimating,smith2014refined} and spatial resolution \cite{qian2013impact,lin2012predictability,smith2014refined} were investigated with GSM and GPS data for smaller populations. Overall, it was found that the predictability increases with temporal resolution and decreases with spatial resolution. The limits of predictability, as defined in \cite{Song2010}, therefore depend on the choice of temporal resolution $\Delta t$  and spatial resolution $\Delta s$ .

Here we make the following conjecture:
\begin{equation}\label{eq:postulate}
\Pi^{(max, min)}(\Delta t, \Delta s) \rightarrow 1 \hspace{0.2cm} \textrm{when} \hspace{0.2cm} \Delta t \rightarrow 0 \textrm{ or } \Delta s \rightarrow \infty. 
\end{equation}
The rationale behind this expression is that the location of the next time bin will almost certainly \emph{not} change in both limits. At small time scales and at large spatial scales you always know where an individual is going to be in the next time bin: he/she will be in the same spatial bin. We therefore argue that the current limits on the predictability of mobility to a large extent reflect stationarity. Previous results therefore mix two different questions, namely
\begin{itemize}
\item How long will an individual stay in his/her current location?
\item Where will he/she go next?
\end{itemize}
Here we propose an analysis that is able to separate out the first question such that we can concentrate on the second. This is achieved by focusing on the \emph{next location}, rather than the location in the next bin. This approach is independent of $\Delta t$, provided a small sampling rate. By introducing a natural length scale, we are able to get a single number for the predictability of human mobility, rather than a function of spatiotemporal resolution. Our new approach shows that the upper limit on the predictability of this type of mobility is around $\sim71\%$, rather than the $>90\%$ found in earlier works. We thereby show that the high upper limits of previous works mostly reflect stationarity, rather than movement.

\section{Data and methods}
\label{sec:1}

\begin{figure*}
  \centering
    \includegraphics[width=\textwidth]{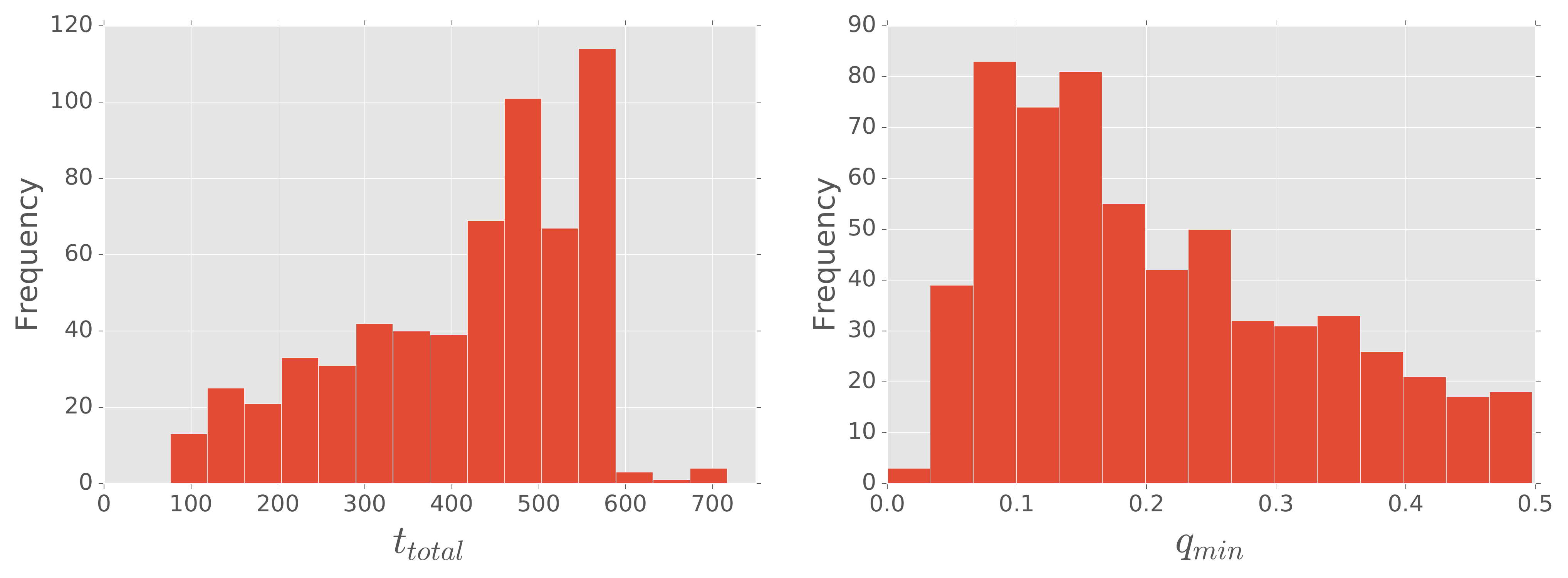}
    \caption{Left panel: the distribution of the number of days that  the participants took part in the experiment. Right panel: the distribution of the fraction of missing data for $\Delta t = 15 $ min.}
	\label{fig:fig1}
\end{figure*}

\paragraph{The Copenhagen Networks Study.} 
Our dataset comes from a large scale study involving approximately 1,000 students over multiple years \cite{stopczynski2014measuring}. Each participant was issued a smartphone capable of recording across multiple channels, including calls, text, bluetooth, and GPS coordinates. In addition to this, the participants answered a questionnaire that, among others, allowed a psychological profile to be inferred. In this paper we mainly use the location data, determined using a combination of the GPS sensors and the network that the phone is connected to. Location data was only available for 849 participants and consists of $\approx 2.4 \cdot 10^8$ data points. The data was collected from February 2012 up to March of 2015, thus covering a multi year span with a substantial fraction participating for more than a year (see left panel of Fig. \ref{fig:fig1}). Each data point consists of latitude and longitude coordinates, together with a timestamp and the accuracy associated with the measurements. These are converted into appropriate time series (see Mobility sequences and predictability for details), and the fraction of bins with unknown locations is denoted $q_{min}$. For our analysis we need $q_{min} \leq 50 \%$ (see Methods). This reduces the number of participants with sufficient data to 604. The right panel of Fig. \ref{fig:fig1} shows the distribution of $q_{min}$ at the lowest temporal scale (15 minutes).

\paragraph{Mobility sequences and predictability.} The raw GPS data needs to be filtered and converted into a history of discrete locations, $T_i$, before the limits of predictability can be determined. This can in principle be done in an infinite number of ways, meaning that the GPS trace from a participant can give rise to many different time series $T_i$ depending on the filtering and mapping chosen. In this work we convert the raw data into two different time series:
\begin{itemize}
\item $T_i^{\textrm{bins}}$: Series of time bins.
\item $T_i^{\textrm{loc}}$: Series of locations.
\end{itemize} 
A detailed description of the filters and mappings are given in the Methods section. 

\begin{figure*}
\centering 
\includegraphics[width=\textwidth]{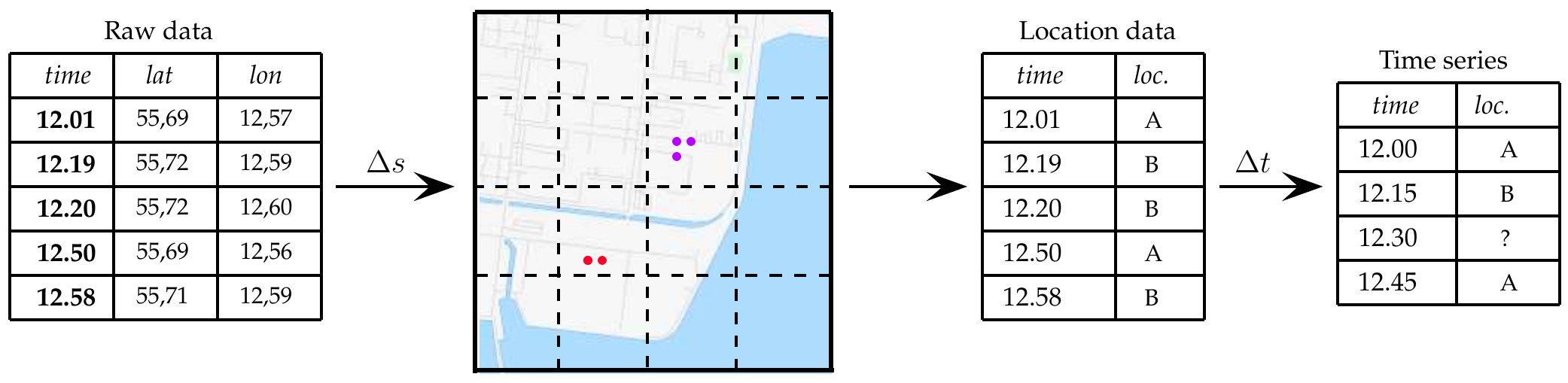}
\caption{Converting the raw data for participant $i$ into a suitable time series, $T_i^{\textrm{bins}}$: After filtering we plot the data points onto a world map overlaid with square grid cells with side lengths $\Delta s$. This converts each data point into a \emph{location} represented by a square grid cell and encoded by a symbol in $T_i^{\textrm{bins}}$. The location data is then resampled such that each bin in $T_i^{\textrm{bins}}$ corresponds to a time interval $\Delta t$. This mobility encoding is similar to that of earlier works and corresponds to the location in a sequence of time bins. We propose to look at a the sequence of locations instead.}
\label{fig:fig2}
\end{figure*}

An illustration of the conversion from GPS-trace into $T_i^{\textrm{bins}}$ is shown schematically in Fig. \ref{fig:fig2}. The two dimensional space is covered by a grid with a grid length given by $\Delta s$. Each square in the grid is represented by a symbol, such that a human trajectory may look like this
\begin{equation}
T_i^{\textrm{bins}} = \left[A,B,B,A,A,A,C..\right] \label{eq:bins}
\end{equation} 
Each symbol corresponds to the grid cell position of a time bin of length $\Delta t$. The construction of this trajectory is equivalent to that of earlier works \cite{jensen2010estimating,Song2010,lin2014mining,qian2013impact,smith2014refined}. As noted earlier, it depends on the spatiotemporal resolution and includes stationarity.

Next we introduce the new mobility encoding $T_i^{\textrm{loc}}$, which aims to describe trajectories by a sequence of unique locations. Details can be found in the Methods section. We start by filtering all the GPS information such that travel between locations is removed. This leaves us with a set of stationary GPS points that are distributed around the preferred places of the individual. We then use a clustering algorithm (DBSCAN \cite{scikit-learn}) on the stationary data points to determine the different locations automatically. This approach results in locations, which better represent the places where individuals spend their time, than the more commonly used Voronoi or square grid cells. 

\begin{figure*}
\centering 
\includegraphics[width=0.55\textwidth]{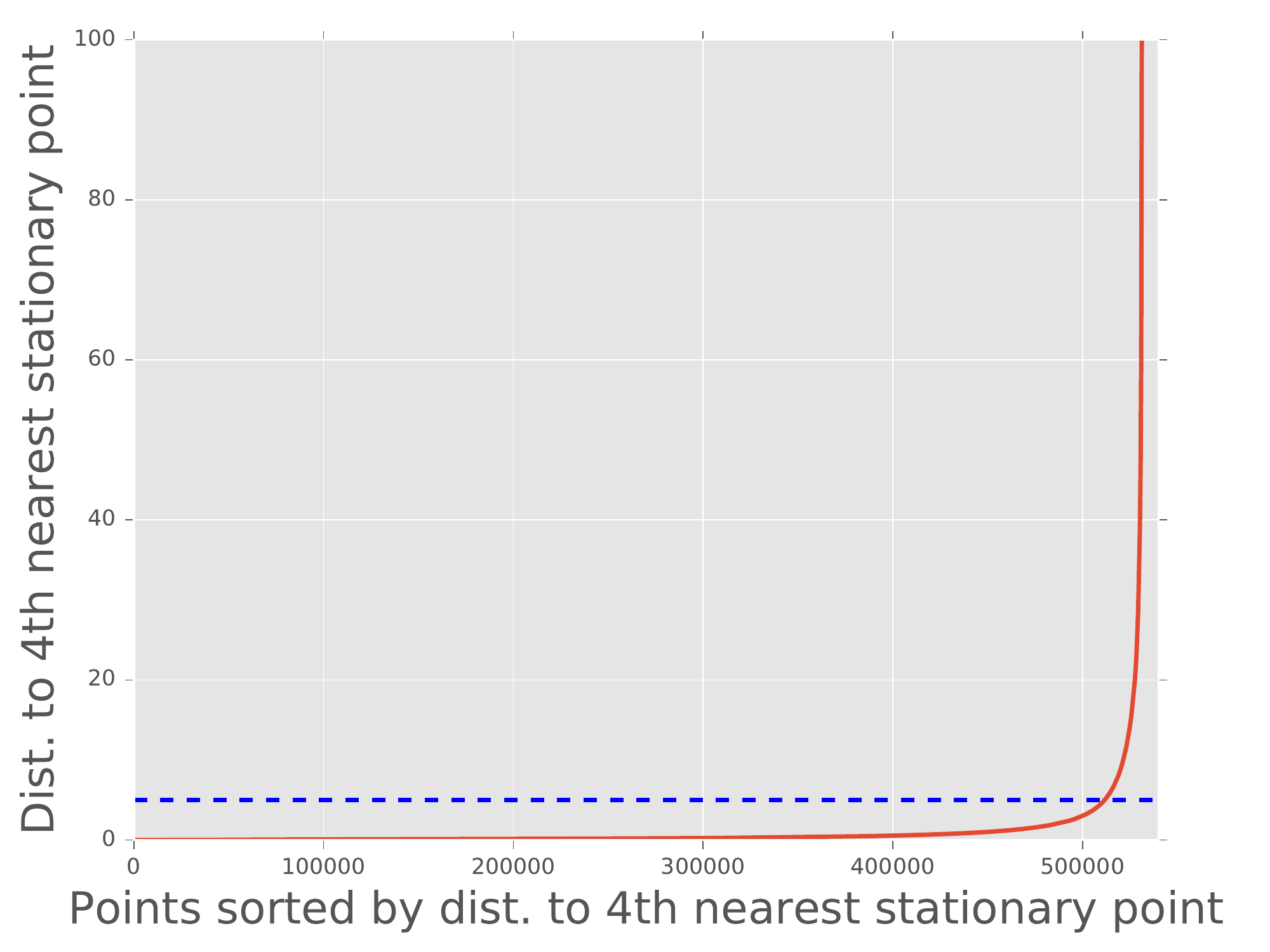}
\caption{The sorted distance to the fourth nearest stationary point is shown for a subset of the entire data set. Standard methods \cite{tan2005dataminig} propose setting the parameter $\epsilon_\textrm{vicinity}$ equal to the distance where there is a sharp increase in the distance to the fourth nearest stationary point. Here we choose $\epsilon=5$ meters as indicated by the dashed line.}
\label{fig:fig3}
\end{figure*}

The clustering algorithm takes a length scale as input, which determines whether or not a stationary data point belongs to a location cluster. Here we use $\epsilon_\textrm{vicinity} = 5 \, \textrm{meters}$ meaning that if a stationary data points is more than five meters from all points in a location cluster, then it is considered as not belonging to that location. This length scale is based on an analysis of "the fourth nearest point"-distribution as proposed in \cite{tan2005dataminig} (see Fig. \ref{fig:fig3}). For the second parameter of the DBSCAN-algorithm, $min\_pts$, we also follow the standards given in the reference, which says to use $min\_pts=4$. This parameter value defines a location cluster as a minimum of four stationary points, i.e. at least 1 hour must be spent in a five meter vicinity during the full sampling period for a cluster to be considered a location.

We can now construct the trajectory of an individual among his/her locations, using the clusters found by the DBSCAN algorithm. In this encoding we do not include the time spent at the different locations, but represent each location by just a single symbol, e.g.:
\begin{equation}
T_i^{\textrm{loc}} = \left[A,B,A,C..\right] \label{eq:locations}
\end{equation}
Compare this with the sequence in (\ref{eq:bins}) and note that the stationarity has been removed, i.e. no similar symbols in a row. 

We expect the sequence of locations to be less predictable than the sequence of time bins, since it encompasses the more complicated spatial dynamics. In order to quantify this intuition, we need a measure of predictability. Here we use a slightly modified version of the scheme developed by \cite{Song2010} (see Methods for details). First, the entropy rate of the mobility sequence is determined using an estimator based on the Lempel-Ziv compression algorithm. Since all the sequences are affected by missing data, one must extrapolate the entropy rate from missing data to full data. By testing our extrapolation on periods with complete data, we find that we can predict the true entropy within 10\%, even when 50\% of the sequence is missing. Having estimated the entropy rate $H_{est}$ we are in a position to determine the upper limit of predictability $\Pi^{max}$. This is done by solving \cite{Song2010}
\begin{align}
H_{est} = &-\Pi^{max} \log_2(\Pi^{max}) \nonumber 
		\\&- (1-\Pi^{max}) \log_2(1-\Pi^{max}) \nonumber 
		\\&+ (1-\Pi^{max}) \log_2 (N-1)
\end{align}
where $N$ is the number of unique locations in the time series. The upper limit found represents a tight upper bound attainable by an appropriate, but for now unknown, algorithm.

We also examine the lower limit of predictability. For the location sequence $T_i^{\textrm{loc}}$, we use a first order Markov chain to predict the next location \cite{lu2013approaching}, i.e. we expect the location that most often follows the current location. If the current location has not been explored before, then we expect the most visited location as the next one. For the time bin sequence $T_i^{\textrm{bins}}$ we use a simple predictor, which expects the current location to continue into the next time bin. This predictor will be referred to as "the trivial predictor" and it measures the amount of stationarity in the mobility sequence.

\section{Results}

\begin{figure*}
  \centering
    \includegraphics[width=\textwidth]{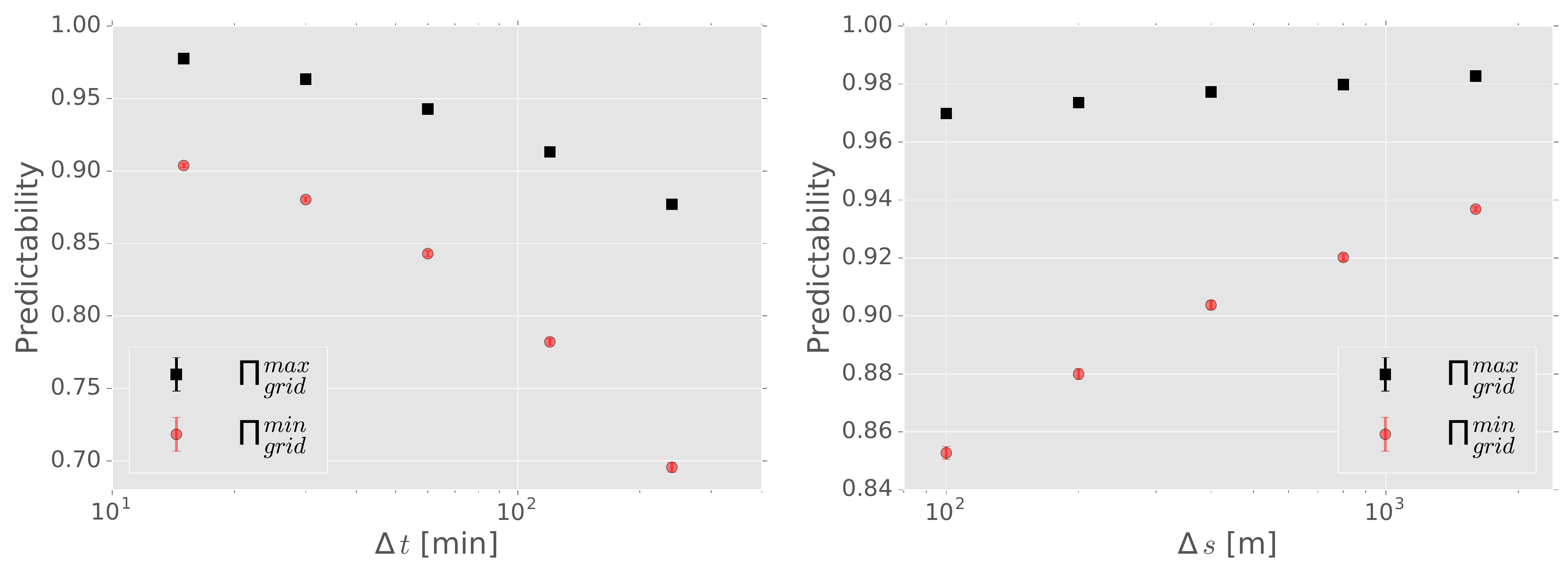}
	\caption{Left panel: The temporal resolution dependency of the upper limits (squares) and lower limits (discs) of predictability for the next bin approach. Each location is represented by a square grid with $\Delta s = 400 $ m. Error bars are included but are smaller than the symbols. Right panel: The spatial dependency at a fixed temporal resolution of $\Delta t = 15 $ min. The lower limit shows that, depending on resolution, 85\% to 94\% of the predictability is due to people \emph{not} moving.}
	\label{fig:fig4}
\end{figure*}

We start by presenting our results for $T_i^{\textrm{bins}}$, i.e. the mobility encoding that people have been using previously. As noted earlier, the predictability of these sequences depend on the spatiotemporal resolution. In the left panel of Fig.~\ref{fig:fig4} we fix $\Delta s = 400 \, \textrm{m}$ and vary $\Delta t$ to determine how the upper and lower limits depend on the temporal scale. The predictability grows towards 1 as the time scale is decreased, just as expected by our conjecture (\ref{eq:postulate}). Note the high performance of the trivial predictor (70\%-91\%). 

Next we fix the temporal scale $\Delta t = 15 \, \textrm{min}$ and vary the spatial scale $\Delta s$ (Fig. \ref{fig:fig4}, right panel). Both the upper limit (squares) and lower limit (discs) increase when $\Delta s$ is increased, again in agreement with (\ref{eq:postulate}). We note that the upper limit is not very sensitive to the spatial scales investigated here ($\Delta s > 100 \textrm{m}$). We furthermore note the impressive performance of the trivial predictor at large spatial scales. For comparison we also compute the limits of predictability at the spatiotemporal scales considered in \cite{Song2010} ($\Delta t = 60 \textrm{min}$ and $\Delta s = 1.7 \textrm{km}$). We find that the trivial predictor is successful in $88.3 \pm 3.8 \%$  of the cases, while the upper bound is $95.5 \pm 1.8 \%$, i.e. almost all of the predictability reflects the fact that people do \emph{not} change location. 

\begin{figure}
\centering 
\includegraphics[width=0.48\textwidth]{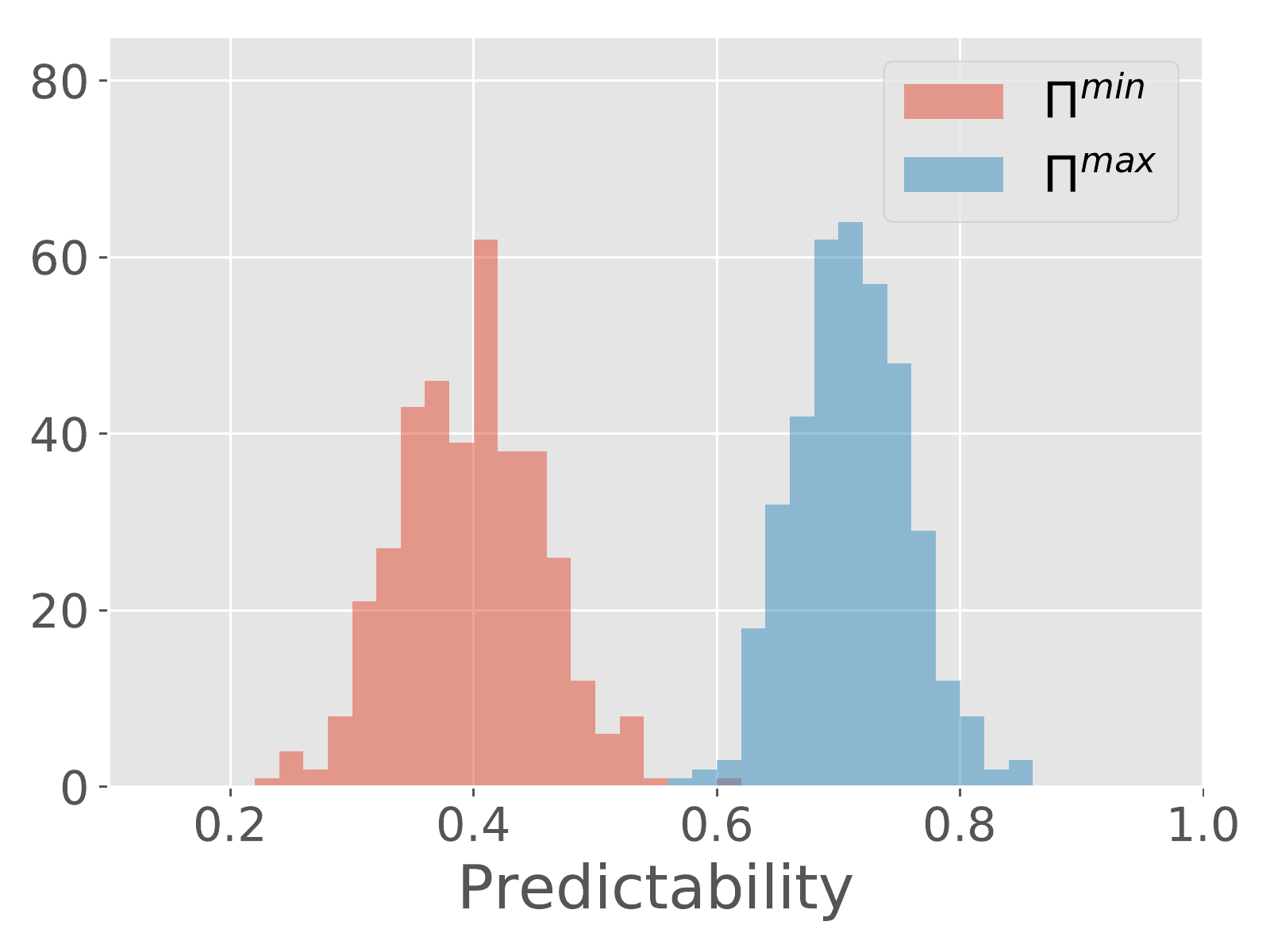}
\caption{The distributions of the upper and lower limits for \emph{next location} predictions. Both results, $\Pi^{max} = 71.1 \pm 4.7 \%$ and $\Pi^{min} = 39.8 \pm 5.9 \%$, are significantly smaller than the limits found for \emph{next bin} predictions (Fig. \ref{fig:fig4}). We conclude that previous work overestimates the predictability of dynamic human mobility.}
\label{fig:fig5}
\end{figure}

The limits presented in Fig. \ref{fig:fig4} follow our postulate and are in agreement with earlier works with smaller populations. We now test what happens when we remove the stationarity from the spatial dynamics, i.e. when we consider the predictability of the next location instead. In Fig. \ref{fig:fig5} we show the distributions of the upper and lower limits for next location predictability. Both limits are strongly reduced when compared to the results for next bin predictability. For the upper limit we find $\Pi^{max} = 71.1 \pm 4.7 \%$, i.e. a significant reduction from the $>90\%$ predictability found in previous works. We find that this value is very robust to increases in the length scale and that it only changes by a few percent as $\epsilon$ is increased towards 25 meters. The lower limit is found to be $\Pi^{min} = 39.8 \pm 5.9 \%$, which is at least $30\%$ lower than any of the lower limits found by the trivial predictor for next bin sequences. 

We note that another group has simultaneously been working on the same data set with the same methods and they have found $\Pi^{max}=0.68$ \cite{cuttone2016understanding}. Despite the close match in results they have actually been using very different DBSCAN paramters, namely $\epsilon_\textrm{vicinity} = 50$ (we use $\epsilon_\textrm{vicinity} = 5$) and $min\_pts=2$ (we use $min\_pts=4$), thereby further underlining the robustness of the results. Our main contribution relative to their work is to derive the length scale from the data, to directly state and investigate conjecture (\ref{eq:postulate}), and to relate the predictability of the next location to psychological factors.

The above results raise the question: what factors impact the predictability of human mobility? Our partial answer to this question can be found in Table \ref{tab:table1}, where we correlate $\Pi^{max}$ to a range of variables. We find that radius of gyration, representing typical distances traveled, does not impact next place predictability. A related result has been reported earlier, using next bin predictability \cite{Song2010}. While this result can seem counterintuitive, our next result is able to shed more light on the matter. $\Pi_{max}$ is anti-correlated with the effective number of places an individual chooses from, when determining where to go next. Therefore, the predictability of an individual does not depend on the reach of his/her travels, but rather on the number of places visited.

Finally, utilizing the psychological profiles of the participants, we are able to examine the impact of their psychological traits on their predictability. The only significant correlation we find here is with extroversion, meaning that the next location of an extroverted individual is statistically harder to predict.

\begin{table}
  \centering
  \caption{Examining which factors impact the predictability of human mobility patterns. $r_g$ is the radius of gyration, eff$_{places}$ is the effective number of places an individual chooses from when changing to a new location and is defined as $2^{H_{unc}}$. We also examine the impact of basic personality traits using the Big Five psychological profile \cite{digman1990personality}. Error bars are determined using the bootstrap method.}
  \label{tab:table1}
  \begin{tabular}{lr}
    \hline
    measure & correlation with $\Pi_{max}$\\
    \hline
    $r_g$                         & $-0.05 \pm 0.05$ \\
    eff$_{places}$                 & $-0.26 \pm 0.05$ \\
    $\Pi_{min}$                   & $0.49 \pm 0.04$ \\
    Agreeableness             & $-0.05 \pm 0.06$ \\
    Conscientiousness         & $ 0.04 \pm 0.06$ \\
    Extroversion              & $-0.13 \pm 0.05$ \\
    Neuroticism               & $ 0.06 \pm 0.06$ \\
    Openess                   & $-0.004 \pm 0.059$ \\
    \hline
  \end{tabular}
\end{table}

\section{Conclusion}
Our results show that it is possible to extract a wide range of upper and lower limits of predictability of human mobility depending on the filtering and discretization scheme chosen. We have shown the strong dependency of "next bin" predictability on spatiotemporal scales. Furthermore, we have shown that the predictability at large spatial scales and small temporal scales mostly reflect stationarity, namely that people stay in the same spatial bin. This raises the need for an alternative approach to estimate the predictability of human mobility patterns. 

The task of predicting human mobility is two fold: how long will a person stay in a certain location and where they will go next. Here we determined an upper limit on the predictability of the latter. We found that the upper limit of this task is much lower than the previously stated ones of $\sim 93\%$. In particular, by using the natural length scale of human locations we found an upper limit on predictability of $71.1 \pm 4.7 \%$. A lower limit was likewise found using a first order Markov chain model with a success rate of $39.8 \pm 5.9 \%$. Overall, our results indicate that it might not be so trivial to predict human mobility after all. 

\section{Methods}

\paragraph{ Converting the raw data into $T_i^{\textrm{bins}}$.} We start by employing an accuracy filter, which removes all the data points with an accuracy below 50 meter. The grid map used is characterized by two parameters: a length scale $\Delta s$ and the origin of the map. The Technical University of Denmark, where most of the participants were enrolled, was chosen as the origin. This ensured that the grid cells had sides of approximately equal length $\Delta s$ at the locations where most of the data was collected. The length scales used are $\Delta s \in \left[ 100, 200, 400, 800, 1600 \right]$ meters. 

Small changes in the origin of the grid map can effect the number of locations detected \cite{lin2012predictability}. To mitigate the possible bias introduced by having a fixed origin of the grid map, we add a random offset for each participant chosen randomly from a uniform distribution on $\left[0, \Delta s\right]$.

Our data was not sampled at a fixed rate. A time binning with a fixed temporal resolution $\Delta t$ allowed us to convert the raw data into a time series. The binning is done such that for each time bin we chose the most visited location. If two or more locations are the most visited locations, then we chose one of them at random. The time scales used are $\Delta t \in \left[15, 30, 60, 120, 240 \right]$ minutes. Time bins with no recorded locations are denoted using a special $?$ marker. Thus we end up with a time series $T_i^{\textrm{bins}}$ which depends primarily on $\Delta s$ and $\Delta t$.

\paragraph{ Converting the raw data into $T_i^{\textrm{loc}}$.} Again we start by employing the accuracy filter. To reduce the number of data points associated with travel, we employ a second filter inspired by the \emph{pause-based} model used in \cite{rhee2011levy}. It detects all the data points which are $15 \pm 1.5$ min apart and for which the distance between the two measurements are less than $100$ m. These two measurements are then averaged into a single data point representing a place where a participant stood still for roughly a quarter of an hour. This filters out most of the travel information in the dataset, except interruptions such as traffic jams and waiting for public transport. 


The list of locations is binned with a fixed temporal resolution $\Delta t = 15$ min as described above. After this we compress every time series such that all instances where a participant stood still for more than one time bin are represented by just a single symbol. This is best explained by an example. A time series obtained by the procedures described above could look like: $T_i = \left[A,?,A,B,B,A,A,A,C..\right]$. After compression this time series is converted into:
\begin{equation}
T_i^{\textrm{loc}} = \left[ A,B,A,C,.. \right]
\end{equation}
The resulting time series are independent of $\Delta t$ provided that $\Delta t$ is small. The smallest sampling rate that we dare use in this study is $\Delta t = 15$, since smaller sampling rates would make it difficult to distinguish stationarity from movement because of the limited accuracy of the GPS. 

\paragraph{ Estimating the entropy rate.} The entropy rate is found using an estimator based on the Lempel-Ziv compression algorithm \cite{Song2010}: 
\begin{equation}\label{eq:estimator}
H_{rate} = \left( \frac{1}{n} \cdot \sum_{i=1}^n \frac{\Lambda_i} {\log(n)} \right)^{-1}
\end{equation}
where $n$ is the length of the time series and $\Lambda_i$ is the length of longest substring in the time series starting from position $i$ and not encountered earlier from position $1$ to $i-1$. This estimator has been shown to converge rapidly towards the entropy rate\cite{kontoyiannis1998nonparametric}. 

The fraction of missing data, $q$, changes the entropy rate estimate. By artificially removing data in complete records we can study possible extrapolation methods. We have used a subset of 47 individuals with a complete location record spanning at least 2 weeks. For each of these complete records we determined $H_{true}$ using the estimator (\ref{eq:estimator}). Removing data from these complete records and comparing the entropy rate determined by our method, $H_{est}$, with $H_{true}$, we found that we could estimate $H_{true}$ within $\pm 10 \%$ as long as $q \leq 0.5$. Our method is thus able to determine the entropy rate even when we only know half of the locations visited. Earlier this method has been used up to $q \leq 0.7$ \cite{Song2010}, but our tests show reliable results only when $q \leq 0.5$. 

Our extrapolation works as follows. For each time series we determine the amount of time the participants location was unknown. This fraction of the total time was denoted $q_{min}$. We then found both $H_{unc}(q)$ and $H_{rate}(q)$ for each $q \in \left[q_{min}, q_{min} +0.05, q_{min} +0.1, .. , 0.9-q_{min} \right]$. Here $H_{unc}$ is the entropy of the time series, found using $H_{unc} = - \sum_{i=1}^N p_i \log_2 (p_i)$, where the sum runs over all the $N$ different locations visited and $p_i$ is the fraction of time spent at $i$. This enabled us to calculate $\sigma(q) = H_{rate}(q) / H_{unc}(q)$. 
Earlier it has been shown \cite{Song2010} that $\sigma(q)$ depends linearly on $q$. This linear relation has not been found when using data with a higher sampling rate \cite{jensen2010estimating}. Our set of complete records showed that $\sigma(q)$ could be fitted well with an offset exponential function. Using these fits we could extrapolate and determine $\sigma_{est} = \sigma(q=0)$. The entropy rate was then found using 
\begin{equation}
H_{est} = \exp^{\sigma_{est}} \cdot H_{unc}(q)
\end{equation}

\section*{List of abbreviations}

\begin{itemize}
\item GSM: Global System for Mobile Communications
\item GPS: Global Positioning System
\item DBSCAN: Density-based spatial clustering of applications with noise
\end{itemize}

\section*{Availability of data and materials}
Data are part of larger study "Social Fabric" involving researchers at the Technical University of Denmark and University of Copenhagen. Due to privacy consideration regarding subjects in our dataset, including European Union regulations and Danish Data Protection Agency rules, we cannot make all data used here publicly available. The data contains detailed information on mobility and daily habits at a high spatio-temporal resolution. We understand and appreciate the need for transparency in research and are ready to make the data available to researchers who meet the criteria for access to confidential data, sign a confidentiality agreement, and agree to work under our supervision in Copenhagen. 

\section*{Ethics approval and consent to participate} 
The "Social Fabric" study was reviewed and approved by the appropriate Danish authority, the Danish Data Protection Agency (Reference number: 2012-41-0664). The Data Protection Agency guarantees that the project abides by Danish law and also considers potential ethical implications. All subjects in the study gave written informed consent.

\section*{Competing interests}
The authors declare that they have no competing interests.

\section*{Author's contributions}
Conceived and designed the study: EI AM. Analyzed the data: EI. Wrote the paper: EI AM. 

\section*{Funding}
The study received funding through the UCPH 2016 Excellence Programme for Interdisciplinary Research. 

\bibliographystyle{spphys}       

\begin{thebibliography}{10}
\providecommand{\url}[1]{{#1}}
\providecommand{\urlprefix}{URL }
\expandafter\ifx\csname urlstyle\endcsname\relax
  \providecommand{\doi}[1]{DOI \discretionary{}{}{}#1}\else
  \providecommand{\doi}{DOI \discretionary{}{}{}\begingroup
  \urlstyle{rm}\Url}\fi

\bibitem{brockmann2006scaling}
D.~Brockmann, L.~Hufnagel, T.~Geisel, Nature \textbf{439}(7075), 462 (2006)

\bibitem{gonzalez2008understanding}
M.C. Gonzalez, C.A. Hidalgo, A.L. Barabasi, Nature \textbf{453}(7196), 779
  (2008)

\bibitem{Song2010}
C.~Song, Z.~Qu, N.~Blumm, A.L. Barab{\'a}si, Science \textbf{327}(5968), 1018
  (2010).
\newblock \doi{10.1126/science.1177170}.
\newblock \urlprefix\url{http://science.sciencemag.org/content/327/5968/1018}

\bibitem{qian2013impact}
W.~Qian, K.G. Stanley, N.D. Osgood, in \emph{Web and Wireless Geographical
  Information Systems} (Springer, 2013), pp. 25--40

\bibitem{rhee2011levy}
I.~Rhee, M.~Shin, S.~Hong, K.~Lee, S.J. Kim, S.~Chong, IEEE/ACM transactions on
  networking (TON) \textbf{19}(3), 630 (2011)

\bibitem{song2010modelling}
C.~Song, T.~Koren, P.~Wang, A.L. Barab{\'a}si, Nature Physics \textbf{6}(10),
  818 (2010)

\bibitem{jiang2016timegeo}
S.~Jiang, Y.~Yang, S.~Gupta, D.~Veneziano, S.~Athavale, M.C. Gonz{\'a}lez,
  Proceedings of the National Academy of Sciences p. 201524261 (2016)

\bibitem{pappalardo2016modelling}
L.~Pappalardo, F.~Simini, arXiv preprint arXiv:1607.05952  (2016)

\bibitem{barbosa2015effect}
H.~Barbosa, F.B. de~Lima-Neto, A.~Evsukoff, R.~Menezes, EPJ Data Science
  \textbf{4}(1), 21 (2015)

\bibitem{pappalardo2015returners}
L.~Pappalardo, F.~Simini, S.~Rinzivillo, D.~Pedreschi, F.~Giannotti, A.L.
  Barab{\'a}si, Nature communications \textbf{6} (2015)

\bibitem{toole2015coupling}
J.L. Toole, C.~Herrera-Yaq{\"u}e, C.M. Schneider, M.C. Gonz{\'a}lez, Journal of
  The Royal Society Interface \textbf{12}(105), 20141128 (2015)

\bibitem{colizza2007modeling}
V.~Colizza, A.~Barrat, M.~Barthelemy, A.J. Valleron, A.~Vespignani, PLoS Med
  \textbf{4}(1), e13 (2007)

\bibitem{kleinberg2007computing}
J.~Kleinberg, Nature \textbf{449}(7160), 287 (2007)

\bibitem{gabaix2003theory}
X.~Gabaix, P.~Gopikrishnan, V.~Plerou, H.E. Stanley, Nature \textbf{423}(6937),
  267 (2003)
  
\bibitem{pappalardo2016analytical}
L.~Pappalardo, M.~Vanhoof, L.~Gabrielli, Z.~Smoreda, D.~Pedreschi,
  F.~Giannotti, International Journal of Data Science and Analytics
  \textbf{2}(1-2), 75 (2016)

\bibitem{frias2012relationship}
V.~Frias-Martinez, J.~Virseda, in \emph{Proceedings of the fifth international
  conference on information and communication technologies and development}
  (ACM, 2012), pp. 76--84

\bibitem{makse1998modeling}
H.A. Makse, J.S. Andrade, M.~Batty, S.~Havlin, H.E. Stanley, et~al., Physical
  Review E \textbf{58}(6), 7054 (1998)

\bibitem{kitamura2000micro}
R.~Kitamura, C.~Chen, R.M. Pendyala, R.~Narayanan, Transportation
  \textbf{27}(1), 25 (2000)

\bibitem{krings2009urban}
G.~Krings, F.~Calabrese, C.~Ratti, V.D. Blondel, Journal of Statistical
  Mechanics: Theory and Experiment \textbf{2009}(07), L07003 (2009)

\bibitem{ranjan2012call}
G.~Ranjan, H.~Zang, Z.L. Zhang, J.~Bolot, ACM SIGMOBILE Mobile Computing and
  Communications Review \textbf{16}(3), 33 (2012)

\bibitem{lin2014mining}
M.~Lin, W.J. Hsu, Pervasive and Mobile Computing \textbf{12}, 1 (2014)

\bibitem{jensen2010estimating}
B.S. Jensen, J.E. Larsen, K.~Jensen, J.~Larsen, L.K. Hansen, in \emph{Machine
  Learning for Signal Processing (MLSP), 2010 IEEE International Workshop on}
  (IEEE, 2010), pp. 196--201

\bibitem{smith2014refined}
G.~Smith, R.~Wieser, J.~Goulding, D.~Barrack, in \emph{Pervasive Computing and
  Communications (PerCom), 2014 IEEE International Conference on} (IEEE, 2014),
  pp. 88--94

\bibitem{lin2012predictability}
M.~Lin, W.J. Hsu, Z.Q. Lee, in \emph{Proceedings of the 2012 ACM Conference on
  Ubiquitous Computing} (ACM, 2012), pp. 381--390

\bibitem{stopczynski2014measuring}
A.~Stopczynski, V.~Sekara, P.~Sapiezynski, A.~Cuttone, M.M. Madsen, J.E.
  Larsen, S.~Lehmann, PloS one \textbf{9}(4), e95978 (2014)

\bibitem{scikit-learn}
F.~Pedregosa, G.~Varoquaux, A.~Gramfort, V.~Michel, B.~Thirion, O.~Grisel,
  M.~Blondel, P.~Prettenhofer, R.~Weiss, V.~Dubourg, J.~Vanderplas, A.~Passos,
  D.~Cournapeau, M.~Brucher, M.~Perrot, E.~Duchesnay, Journal of Machine
  Learning Research \textbf{12}, 2825 (2011)

\bibitem{tan2005dataminig}
P.N. Tan, M.~Steinbach, V.~Kumar, \emph{Introduction to Data Mining} (Pearson,
  2005)

\bibitem{lu2013approaching}
X.~Lu, E.~Wetter, N.~Bharti, A.J. Tatem, L.~Bengtsson, Scientific reports
  \textbf{3} (2013)

\bibitem{cuttone2016understanding}
A.~Cuttone, S.~Lehmann, M.C. Gonz{\'a}lez, arXiv preprint arXiv:1608.01939
  (2016)

\bibitem{digman1990personality}
J.M. Digman, Annual review of psychology \textbf{41}(1), 417 (1990)

\bibitem{kontoyiannis1998nonparametric}
I.~Kontoyiannis, P.H. Algoet, Y.M. Suhov, A.J. Wyner, Information Theory, IEEE
  Transactions on \textbf{44}(3), 1319 (1998)


\bibitem{jurdak2015understanding}
R.~Jurdak, K.~Zhao, J.~Liu, M.~AbouJaoude, M.~Cameron, D.~Newth, PloS one \textbf{10}(7), e0131469 (2015)

\end{thebibliography}

\end{document}